# Intrinsic Origin of Enhancement of Ferroelectricity in SnTe Ultrathin Films

Kai Liu[1,2], Jinlian Lu[1,2], Silvia Picozzi[3], Laurent Bellaiche[4]*, and Hongjun Xiang[1,2]*

[1]*Key Laboratory of Computational Physical Sciences (Ministry of Education), State Key Laboratory of Surface Physics, and Department of Physics, Fudan University, Shanghai 200433, P. R. China*

[2]*Collaborative Innovation Center of Advanced Microstructures, Nanjing 210093, P. R. China*

[3]*Consiglio Nazionale delle Ricerche CNR-SPIN Via dei Vestini 31, Chieti 66100, Italy*

[4]*Physics Department and Institute for Nanoscience and Engineering University of Arkansas, Fayetteville, Arkansas 72701, USA*

E-mail: laurent@uark.edu, hxiang@fudan.edu.cn

**Abstract**

Previous studies showed that, as ferroelectric films become thinner, their Curie temperature ($T_c$) and polarization below $T_c$ both typically decrease. In contrast, a recent experiment [Chang *et al.*, Science **353**, 274 (2016)] observed that atomic-thick SnTe films have a higher $T_c$ than their bulk counterpart, which was attributed to *extrinsic* effects. Here, we find, using first-principles calculations, that the 0K energy barrier for the polarization switching (which is a quantity directly related to $T_c$) is higher in most investigated *defect-free* SnTe ultrathin films than that in bulk SnTe, and that the 5-unit-cell (UC) SnTe thin film has the largest energy barrier as a result of an interplay between hybridization interactions and Pauli repulsions. Further simulations, employing a presently developed effective Hamiltonian, confirm that free-standing defect-free SnTe thin films have a higher $T_c$ than bulk SnTe, except for the 1-UC case. Our work therefore demonstrates the possibility to *intrinsically* enhance ferroelectricity of ultrathin films by reducing their thickness.

Ferroelectric (FE) materials in which a spontaneous polarization can be switchable by an electric field have wide applications such as high-density nonvolatile memories [1-3]. The continuous demand for device miniaturization has resulted in the increased demand of nanometric FE thin films [2,4,5]. Nanoscale ferroelectrics are also fascinating from a fundamental point of view, since their properties can be dramatically different from that in of bulk ferroelectrics [6,7].

A suppression of the out-of-plane polarization is known to occur in thin films under open-circuit-like electrical boundary condition, because of depolarization field effects [8]. Regarding the in-plane component of the polarization, finite-size scaling theory predicts that FE Curie temperature ($T_c$) shifts to lower temperatures as compared to the bulk value, as FE films become thinner [9,10]. Such features are consistent with measurements and computations demonstrating that, typically, the FE $T_c$ becomes lower, and, as a result, the electrical polarization becomes smaller for any temperature below $T_c$, as the thickness of FE thin films is reduced [8,11-15]. It is therefore rather surprising that Chang *et al.* recently observed that the $T_c$ of atomic-thick SnTe is *higher* than that of SnTe bulk [16]. In order to reconcile such observation with the aforementioned features, this unusual phenomenon was mainly explained in terms of an *extrinsic* effect, namely that there are less Sn vacancies and lower free carrier density in SnTe thin films. It is, however, interesting to investigate how $T_c$ *intrinsically* changes with the thickness in perfectly defect-free SnTe thin films. In particular, is this currently unknown dependency in line with the common belief that $T_c$ should decrease as the films become thinner, or are there any new effects awaiting to be revealed in SnTe thin films leading to an *intrinsic* enhancement of their ferroelectricity when their thickness is reduced?

In this Letter, we carry out first-principles calculations to determine such intrinsic dependency. We find that the 0K energy barrier for the polarization switching in SnTe thin films is higher than that in bulk SnTe when the thickness is larger than 2-unit cell (UC), and that the 5-UC SnTe thin films has the largest energy barrier. Such results strongly suggest that $T_c$ of SnTe thin films is not only larger than that of bulk SnTe but also reaches its maximum for the 5-UC thickness. These unusual phenomena are found to originate from the subtle interplay between hybridization interactions (HIs), which are essential for stabilizing ferroelectricity, and Pauli repulsions (PRs), that tend to suppress ferroelectricity. More precisely, (i) the increase of energy barrier when decreasing the thickness from 13-UC to 5-UC, is due to the fact that the surface Sn atoms have weaker PRs and thus a smaller force constant than the inner Sn atoms; and (ii) the decrease of energy barrier from 5-UC to 1-UC arises from the concomitant decrease of HIs. We also developed an effective Hamiltonian for free-standing defect-free SnTe thin films under open-circuit electrical boundary conditions, which does confirm that SnTe films intrinsically have a higher $T_c$ than bulk SnTe, except for the 1-UC case. Our work therefore suggests that, unlike commonly believed, it is possible to intrinsically enhance ferroelectricity by reducing the thickness of ultrathin films.

Let us first recall that bulk SnTe is a narrow-gap (~0.2 eV) semiconductor that possesses the rock-salt structure [17]. At the FE $T_c$ (98 K in a sample with low carrier density [18]), bulk SnTe goes through a cubic paraelectric (PE) to rhombohedral FE phase transition, and the two sublattices of Sn and Te atoms are displaced from each other along the [111] direction -- giving rise to a spontaneous polarization along that direction [19]. On the other hand, in SnTe ultrathin films being under open-circuit electrical boundary conditions, the polarization is along the in-plane [110] direction because the depolarization field annihilates the out-of-plane polarization [16,20-22]. Here, we perform density function theory (DFT) calculations [see Section 1 of Supplemental Material (SM) for computational details] to relax the structures of the FE phase. Regarding the centrosymmetric PE phase of thin films, we optimize the slab cut from the cubic bulk SnTe but keeping the same symmetry. The PE and FE structures of 1-UC and 2-UC SnTe thin films are shown in Fig. 1. For the 1-UC SnTe thin film, all $Sn^{2+}$ ions move along the [110] direction so that each $Sn^{2+}$ ion is three-fold coordinated. The resulting FE phase with two puckered atomic layers is similar to monolayer black phosphorus [23]. The optimized in-plane lattice constants are slightly anisotropic ($a$ = 4.58 Å and $b$ = 4.55 Å) [16]. The multilayer FE SnTe thin films can be regarded as a stacking of 1-UC FE SnTe films along the $c$-axis [see Fig.

1(b) for the case of 2-UC thin film].

Since the magnitude of the polarization (see Section 2 of SM) is related to ionic displacements, we plot in Fig. 1(e) the FE displacements of $Sn^{2+}$ ions at different layers for different SnTe thin films (1-UC, 2-UC, 5-UC, and 10-UC). The FE displacements of $Sn^{2+}$ ions are computed by assuming that the positions of the $Te^{2-}$ ions are fixed to those of the PE phase. Note that each SnTe unit-cell contains a double layer. Several interesting trends can be seen: i) For a SnTe thin film with a given thickness, the surface $Sn^{2+}$ ions displace more than the inner $Sn^{2+}$ ions; ii) For very thin films (i.e., 1-UC and 2-UC), FE displacements become smaller as the films become thinner; and iii) For multilayer SnTe thin films (i.e., 5-UC or 10-UC), the FE displacements of $Sn^{2+}$ ions display an odd-even oscillating behavior with respect to the layer number, e.g., the motion of the $Sn^{2+}$ ion of the second layer is smaller than that of both the first and third layers.

To study the stability of ferroelectricity in SnTe thin films, we calculate the energy difference between the PE and FE phases (i.e., the energy barrier between two FE states with opposite polarizations) of SnTe thin films as a function of thickness (see the red line in Fig. 2). One can see that this barrier first increases with the thickness from 1-UC to 5-UC, and decreases as the thickness increases for films thicker than 5-UC. Strikingly, the energy barriers of SnTe thin films from 2-UC to 13-UC are higher than that of bulk SnTe, which strongly suggests that defect-free SnTe thin films have a higher Curie temperature than bulk SnTe with no defects (as we will confirm later) -- in contrast to the previous belief that ferroelectricity in thin films is reduced with respect to the bulk case.

To check the effect of strain on the dependence of energy barrier on the film thickness, we also computed the energy barriers in the case where the in-plane lattice constants of the PE and FE phases are fixed to those of bulk SnTe ($a = 4.5254$ Å). The energy barrier also has a maximum for 5-UC [see Fig. 2(b)], implying that strain does not qualitatively affect the dependence of energy barrier on film thickness. Therefore, the interactions related to Sn displacements are responsible for the intriguing thickness dependency of the energy barrier. To investigate the interactions related to

Sn displacements in more details, we determined the layer force constant $k_{DFT}^i$ (see Ref. [24] for more details) by computing and fitting the total energy $E$ as a function of the Sn displacement $x$ in the $i$-th layer by the form $E = \frac{1}{2} k_{DFT}^i x^2 + E_{PE}$ (where $E_{PE}$ is the total energy of the PE phase). The chosen displacement $x$ is so small (e.g., ~0.01 Å) that fourth and higher order terms can be neglected here. The layer force constants for 1-UC, 2-UC, 5-UC and 10-UC SnTe thin films are plotted in Fig. 3. The $Sn^{2+}$ ions of the surface layer have a smaller force constant than the $Sn^{2+}$ ions of the inner layers for multilayer thin films. Moreover, the layer force constant of the surface layer basically increases as the films become thinner, as shown in the inset of Fig. 3. In particular, the layer force constants of the surface layer for films thinner than 3-UC (see also Fig. S4) is significantly larger than those in thicker films. For a given thin film (e.g., 5-UC and 10-UC thin films), the layer force constants display an oscillating odd-even behavior with respect to the layer number. The dependence of the layer force constants is in line with the dependence of the FE displacements [see Fig. 1(e)] since a smaller force constant indicates a stronger FE instability [25]. The dependence of the layer force constants can also explain the dependence of the energy barrier on the film thickness. In fact, why the energy barrier does not always increase with thickness can be understood as follows: since the $Sn^{2+}$ ions of the surface layer have a smaller force constant than the $Sn^{2+}$ ions of the inner layer, thinner films tend to display stronger ferroelectricity (higher energy barrier) as they have a larger surface-to-volume ratio. Furthermore, the reason why 5-UC SnTe thin film has the highest energy barrier is that the layer force constants in the thinnest films (i.e., 1-UC and 2-UC) are in average much larger than those in films thicker than 3-UC, as shown in Fig. 3.

Now we attempt to understand the microscopic mechanisms for the unusual behavior of the layer force constants. It is well-known that ferroelectricity in bulk SnTe is mainly caused by the hybridization between the empty 5p orbitals of the lone-pair $Sn^{2+}$ ion and the occupied 5p orbitals of the $Te^{2-}$ ion, i.e., second-order Jahn-Teller effects [26,27]. To see the effect of orbital hybridization on the layer force constants, we computed these constants by using the band energies from the Tight-Binding (TB) simulations (see Section 1 of SM). As shown in Fig. 4(a), the layer force constants contributed by the HIs are found to be negative, suggesting that HIs favor FE displacements, in agreement with a previous study [28]. The oscillating odd-even behavior of the dependence of the HIs-related layer force constants and the fact that the second layer (i.e., subsurface) has the largest force constant are in line with the dependence of the total layer force constants from the DFT calculations [compare Figs. 3 and 4(a)]. The HIs-related layer force constants in 1-UC and 2-UC thin films are much larger than those in thicker films, most likely because the much larger band gaps arising from the quantum confinement in ultrathin films weaken the HIs. If considering only the HIs-related layer force constants, ferroelectricity in thinner films would always be weaker than that in thicker films since thinner films have the largest averaged HIs-related layer force constants. Therefore, although the consideration of sole HIs can explain some of the trends of the energy barrier, it cannot solely provide an explanation of why the 5-UC thin film has the highest energy barrier.

To resolve this issue, it is important to realize that the off-centric FE displacements arise due to the delicate balance between the HIs (which favors ferroelectricity [28-30]) and the Pauli repulsions (PRs) (which tends to keep systems centrosymmetric [31,32,33]). We will thus now see how PR affects the layer force constants. Due to the acoustic sum rule, we can decompose the layer force constant of the $Sn^{2+}$ ions of the $i$-th layer ($k_{DFT}^i$) into intralayer ($k_{intra}^i$) and interlayer ($k_{inter}^i$) contributions (see Ref. [24] for details). Note that all the PRs within the $i$-th layer are reflected in the intralayer force constant $k_{intra}^i$. As an example, we show the intralayer and interlayer contributions of the layer force constants for the surface and inner layers of the 5-UC SnTe thin film in left part of Fig. 4(b). We find that the interlayer contributions for both the surface and inner layers are positive and of similar magnitude. The large difference in the layer force constant between the surface and inner layers arises because of the much larger positive intralayer contribution in the inner layer case. The intralayer contribution $k_{intra}^i$ can be seen as a sum of the HI-related contribution and the PR-related contribution between the $Sn^{2+}$ ion and the in-plane $Te^{2-}$ ion. Since the HI-related contribution to $k_{intra}^i$ of the surface layer is not less than that of the inner layer, the much larger positive intralayer contribution in the inner layer case must be due to the fact that the PR-related contribution to $k_{intra}^i$ of the surface layer is more negative than that of the inner layer. Actually, this can be understood by a simple ionic radius argument: the surface $Sn^{2+}$ and $Te^{2-}$ ions in the PE SnTe thin film are five-fold coordinated while the inner $Sn^{2+}$ and $Te^{2-}$ ions are six-fold coordinated, and it is well known that the radius of an ion with smaller coordination number is smaller than that with a larger coordination number. Since the Sn-Te bond length of the surface layer is similar to that of the inner layer in the PE phase, it is expected that the Sn-Te bond of the surface layer can be compressed much more easily than that of the inner layer. Our argument is further supported by considering a hypothetical 5-UC SnTe thin film being under 7.2% tensile strain. In this case, the layer force constants of the surface and inner layers become negative [(see right part of Fig. 4(b)], suggesting that the FE instability is enhanced, in agreement with the fact that tensile strain strengthens in-plane ferroelectricity. Interestingly, the layer force constants of the inner layer is now close to that of the surface layer. This is because the Sn-Te PR is greatly reduced, as the Sn-Te bond length is now larger than the sum of the $Sn^{2+}$ and $Te^{2-}$ ionic radii. Our above argument that there is less PR in a surface layer should be generally applicable to other systems. In fact, this holds even in a non-FE ionic system (see Section 4 of SM for the result on MgO).

For comparison, we also examine ferroelectricity in thin films made of other IV–VI compounds (i.e., GeTe and PbTe). For the GeTe thin films, the energy barrier increases with the film thickness, i.e., there is no maximum in the energy barrier curve (see Section 5 of SM). This can be understood since the HI in GeTe is so strong that the PR contribution to the layer force constant is less important. The intriguing behavior that a thinner film may possess a stronger ferroelectricity occurs when the HI and PR are of comparable magnitude. In fact, if we reduce the PR in PbTe thin films by applying a tensile strain, we also observe a maximum in the energy barrier curve (see Section 6 of SM). Furthermore, we demonstrate that this mechanism is generally applicable to other systems. For example, we find that the energy barriers and polarizations of $TiO_2$-terminated $SrTiO_3$ [001] thin films under a small tensile strain (lateral lattice constants $a = b = 4.0955$ Å) decrease with the film thickness (see Section 7 of SM).

Finally, we estimate the $T_c$ of SnTe thin films by developing an effective Hamiltonian and performing parallel tempering Monte Carlo (PTMC) simulations for these 2D systems. As shown in Fig. S9, $T_c$ increases from 1-UC to 5-UC and then decreases when further increasing the film thickness. The tendency of $T_c$ with thickness is in line with that of the energy barrier. Our simulations show that defect-free bulk SnTe has a lower $T_c$ (that is, 38 K) than most of SnTe thin films (i.e., $T_c$ for 2-UC thin film is 47 K). However, the defect-free 1-UC SnTe film has a lower $T_c$ (namely, 30 K) than defect-free bulk SnTe, in contrast with the experimental result [16]. This suggests that the high-$T_c$ measured for the 1-UC SnTe film [16] is partly due to extrinsic effects (e.g., defects [37,38] (see Section 10 of SM), van der Waals interactions between the SnTe thin film and the substrate, charge transfer between the SnTe thin film and the substrate, etc.). Although $T_c$ is underestimated in our simulations for bulk SnTe and SnTe thin films, the qualitative trend from the simulations should be correct. In fact, with more accurate (and more demanding) methods that predict a larger energy barrier and thus a higher $T_c$, the qualitative dependence of the energy barriers on the film thickness remains unchanged (see Section 9 of SM). We also perform additional PTMC simulations on a model Hamiltonian to prove that the smaller force constant for the surface ion is indeed the key to the non-monotonic behavior of $T_c$ as a function of the thickness (see Section 11 of SM).

In summary, based on first-principles calculations and effective Hamiltonian simulations, we revealed that the FE switching energy barrier and $T_c$ in free-standing defect-free SnTe thin films first increase with thickness when the film thickness is less than 5-UC, and then decreases with thickness for thicker films. These atypical behaviors originate from a subtle interplay between HIs and PRs. Our work is thus promising towards the realization of miniaturized FE devices utilizing ultrathin films.

**Acknowledgments**

Work at Fudan is supported by NSFC (11374056), the Special Funds for Major State Basic Research (2015CB921700), Program for Professor of Special Appointment (Eastern Scholar), Qing Nian Ba Jian Program, and Fok Ying Tung Education Foundation. H.X. also thank Dr. Junwei Liu for useful discussions. L.B. acknowledges the ONR Grants N00014-12-1-1034 and N00014-17-1-2818 and ARO Grant No. W911NF-16-1-0227.

[24] To compute the layer force constant $k^i_{\text{DFT}}$ of the $Sn^{2+}$ ions of the $i$-th layer, we can displace the $Sn^{2+}$ ions of the $i$-th layer along the $x$-axis by $\delta$ and calculate the force $F^{Sn}_{ix}$ exerted on it with DFT. One then gets $k^i_{\text{DFT}} = -\frac{F^{Sn}_{ix}}{\delta}$. One can further decompose the layer force constant into an intralayer contribution and an interlayer contribution using the acoustic sum rule. Due to translational invariance, the sum of all the forces on all atoms should vanish: $\sum_j (F^{Sn}_{jx} + F^{Te}_{jx}) = 0$, where $j$ is the layer-index. Thus, $k^i_{\text{DFT}} = k^i_{\text{intra}} + k^i_{\text{inter}}$, where the intralayer contribution is $k^i_{\text{intra}} = \frac{F^{Te}_{ix}}{\delta}$

and the interlayer contribution is $k^i_{\text{inter}} = \frac{1}{\delta} \sum_{j \neq i} (F^{\text{Sn}}_{jx} + F^{\text{Te}}_{jx})$.

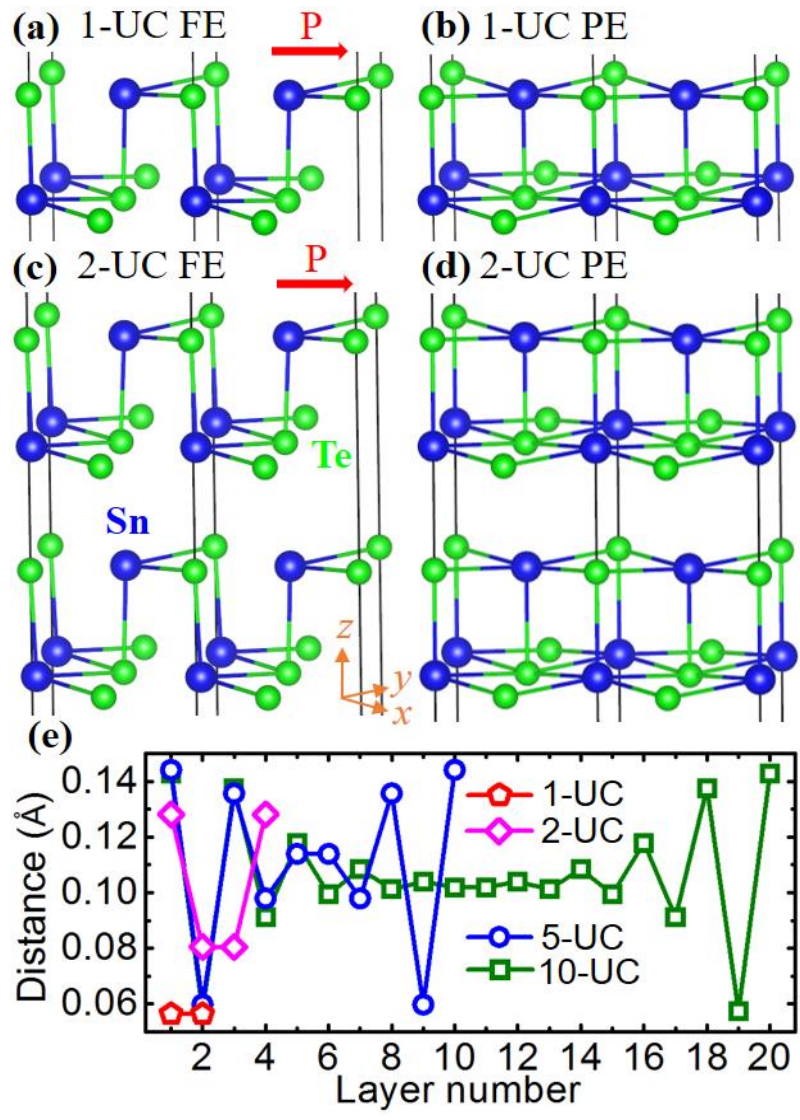

Figure 1. Crystal structures of SnTe thin films. Ferroelectric (FE) phases of (a) 1-UC and (c) 2-UC SnTe thin films. The corresponding paraelectric phases are shown in (b) (d). The red arrows indicate that polarizations are along the [110] direction. (e) The FE displacements of $Sn^{2+}$ ions at different layers as a function of the layer number.

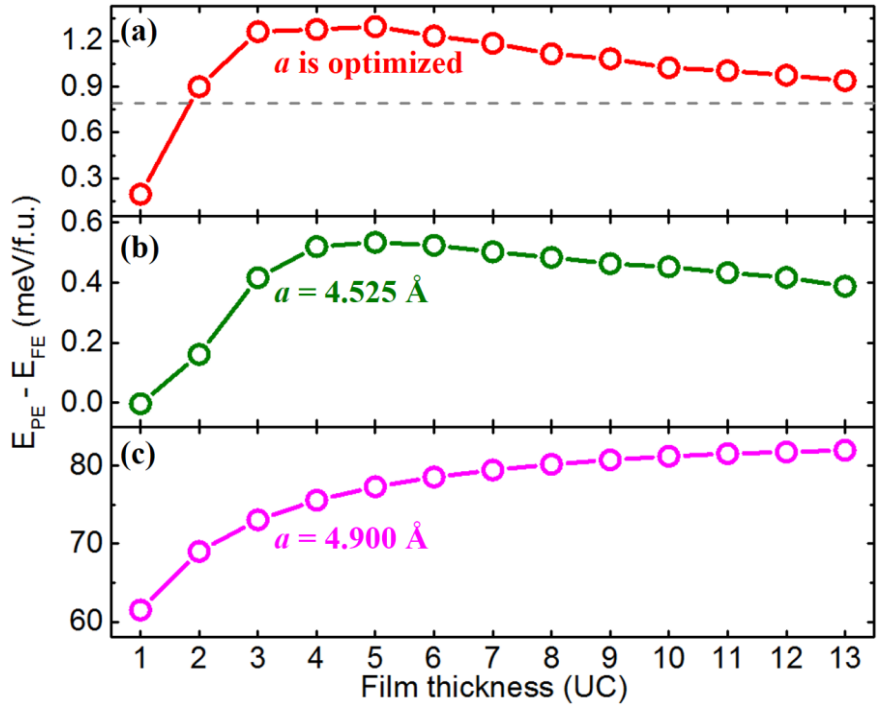

Figure 2. Energy differences between the PE and FE phases of SnTe films. (a) The lateral lattice constants $a$ and $b$ are fully optimized; (b) lattice constants $a$ and $b$ are fixed to the bulk lattice constant of 4.525 Å; and (c) lattice constants a=b=4.900 (a 7.2% tensile strain). The horizontal dashed line shows the result of bulk SnTe.

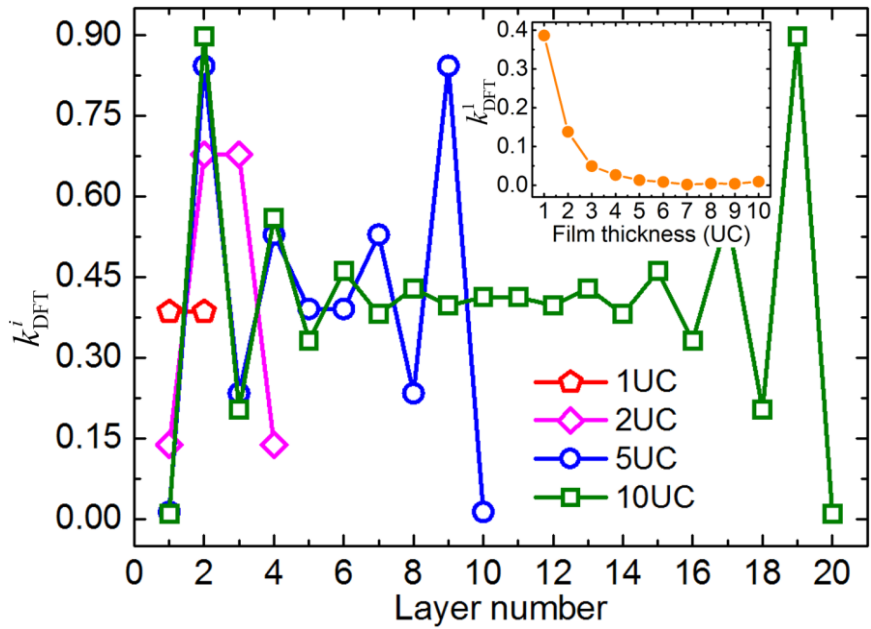

Figure 3. Layer force constants of $Sn^{2+}$ ions as a function of the layer number, as fitted with the DFT total energies. The insert displays the force constant of the surface $Sn^{2+}$ ion as a function of film thickness.

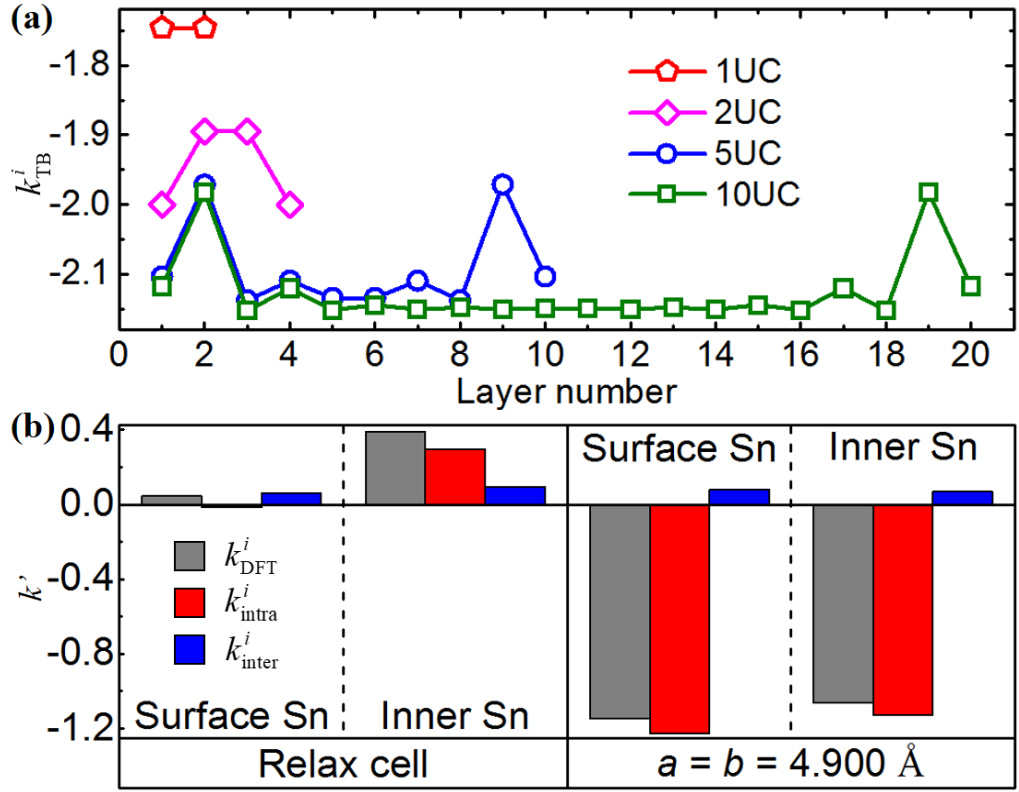

Figure 4. (a) Layer force constants of $Sn^{2+}$ ions as a function of layer number, as fitted with the TB band energies. (b) Decomposition of layer force constants $k^i_{DFT}$ into $k^i_{intra}$ and $k^i_{inter}$ for the 5-UC SnTe thin film (see main text).

*Supplemental Material for*

**Intrinsic Origin of Enhancement of Ferroelectricity in SnTe Ultrathin Films**

Kai Liu[1,2], Jinlian Lu[1,2], Silvia Picozzi[3], Laurent Bellaiche[4]*, and Hongjun Xiang[1,2]*

*[1]Key Laboratory of Computational Physical Sciences (Ministry of Education), State Key Laboratory of Surface Physics, and Department of Physics, Fudan University, Shanghai 200433, P. R. China*

*[2]Collaborative Innovation Center of Advanced Microstructures, Nanjing 210093, P. R. China*

*[3]Consiglio Nazionale delle Ricerche CNR-SPIN Via dei Vestini 31, Chieti 66100, Italy*

*[4]Physics Department and Institute for Nanoscience and Engineering University of Arkansas, Fayetteville, Arkansas 72701, USA*

E-mail: laurent@uark.edu, hxiang@fudan.edu.cn

## 1. Computational details

**DFT calculations.** Our first-principles density functional theory (DFT) calculations are performed on the basis of the projector augmented-wave method [1] encoded in the Vienna *ab-initio* simulation package (VASP) [2,3]. We explicitly consider 14 and 6 valence electrons for Sn and Te, respectively. To be more specific, we take into account the ten 4d electrons of Sn as valence electrons. This procedure is referred as "Sn_d" hereafter. For comparison, we also adopt other procedures, as we will discuss later. We employ the generalized-gradient approximation (GGA) of Perdew *et al.* [4]. The plane-wave cutoff energy is set to be 301.4 eV. The crystal structures are considered to be fully optimized when the ionic forces are less than 0.01 eV/Å. A 12×12×1 (12×12×12) Monkhorst-Pack k-mesh is employed to sample the Brillouin zone of all SnTe thin films (bulk SnTe). We use 21 data points to fit the energy curve as a function of the displacement of $Sn^{2+}$ ion of the *i*-th layer ($E = \frac{1}{2}k_{DFT}^{i}x^2 + E_{PE}$) so as to obtain the layer force constant $k_{DFT}^{i}$ of the *i*-th layer.

**Tight-binding simulations.** We estimate the layer force constant originating from the hybridization interactions by performing tight-binding (TB) simulations, for which the Hamiltonian includes the on-site term and hopping term between neighboring Sn and Te ions:

$$H = \sum_{i,\alpha}(\varepsilon_{Sn}c^+_{Sn,i,\alpha}c_{Sn,i,\alpha} + \varepsilon_{Te}c^+_{Te,i,\alpha}c_{Te,i,\alpha}) + \sum_{\langle Sn,i,Te,j\rangle,\alpha}[t_{ij\alpha}(c^+_{Sn,i,\alpha}c_{Te,j,\alpha} + h.c.)],$$

where $\varepsilon_{Sn}$ and $\varepsilon_{Te}$ are the on-site energies for the Sn 5p and Te 5p orbitals, respectively, and $i$ and $\alpha$ denote the site and orbital ($p_x$, $p_y$, $p_z$) index, respectively. Note that $s$ orbitals of Sn and Te ions are neglected here as they will not to change the main results. The second sum over $\langle Sn,i,Te,j\rangle$ represents the neighboring pairs between the Sn ion at $i$ and the Te ion at $j$. The hopping parameter $t_{ij\alpha}$ is computed using the Slater-Koster interaction parameters $t_{pp\sigma}$ and $t_{pp\pi}$ [5]. To estimate the Slater-Koster interaction parameters $t_{pp\sigma}$ and $t_{pp\pi}$, we constructed the maximally localized Wannier functions of cubic bulk SnTe with the Wannier90 program [6,7].

To calculate the contribution of hybridization interactions to layer force constant $k^i_{TB}$ for the Sn ions in the $i$-th layer of the SnTe thin film, we computed the TB total energy, $E^{PE}_{TB}$, as a function of the displacement $x$ of the Sn ions in the $i$-th layer. In the TB simulation, the total energy is taken to be the band energy, i.e., the sum of occupied eigenvalues of the TB Hamiltonian. We then fitted the TB total energy by the function $E_{TB} = \frac{1}{2}k^i_{TB}x^2 + E^{PE}_{TB}$ to obtain $k^i_{TB}$ ($E^{PE}_{TB}$ is the TB total energy of the paraelectric phase of the considered SnTe thin film).

**Monte-Carlo simulations with an effective Hamiltonian.** To estimate the FE Curie temperature of SnTe thin films, we performed parallel tempering Monte Carlo (PTMC) simulations [8,9] with an effective Hamiltonian. Note that the effective Hamiltonian schemes were widely and successfully used to study finite-temperature properties of FE perovskite oxides [10,11], but the applications of effective

Hamiltonians to other structures are rare. Following the seminar work by Zhong *et al.* [10], the effective Hamiltonian employed in this work contains five parts: a local-mode self-energy, a long-range dipole-dipole interaction, a short-range interaction between soft modes, an elastic energy, and an interaction between the local modes and local strain. To be more specific,

$$E^{tot} = E^{self}(\{\mathbf{u}\}) + E^{dpl}(\{\mathbf{u}\}) + E^{short}(\{\mathbf{u}\}) + E^{elas}(\{\eta_l\}) + E^{int}(\{\mathbf{u}\},\{\eta_l\}),$$

where $\mathbf{u}$ represents the local soft mode (displacements of Sn ions for SnTe systems). For SnTe thin films, the local soft modes include the in-plane displacements of Sn ions, while, for bulk SnTe, the local soft modes include displacements of Sn ions along all three Cartesian directions. For SnTe thin films, we only consider the three strain components ($\eta_l = \eta_1, \eta_2,$ or $\eta_6$) in Voigt notation that are related to the in-plane lattice constants [note that we, e.g., numerically found that the thickness along the c direction of the paraelectric (PE) and FE phases of the 5-UC film are very close to each other (namely, 28.9006 Å and 28.8733 Å, respectively), which explains why we did not incorporate the out-of-plane component of the strain for the films]. On the other hand, for bulk SnTe, all six components of the strain tensor were taken into account.

For the self-energy part, $E^{self}(\{\mathbf{u}\})$, we included all the self-energy terms up to fourth order of $\mathbf{u}$. In our simulations, we do not separate the contribution from the dipole-dipole interaction $E^{dpl}(\{\mathbf{u}\})$ and the short-range interaction $E^{short}(\{\mathbf{u}\})$. Instead, we find it more convenient to adopt a second-order pair interaction $E^{pair}(\{\mathbf{u}\}) = \frac{1}{2}\sum_{i \neq j}\sum_{\alpha\beta} J_{ij,\alpha\beta} u_{i\alpha} u_{j\beta}$. The pair interactions include all Sn-Sn pairs being distant by less than 14.8 Å. The elastic energy $E^{elas}(\{\eta_l\})$ is computed with the elastic constants of the PE structure. To describe the coupling between strain and local modes, we considered the dependence of the second-order on-site interaction on the strain, namely, $E^{int}(\{\mathbf{u}\},\{\eta_l\}) = \frac{1}{2}\sum_i \sum_{l\alpha\beta} B_{l\alpha\beta} \eta_l u_\alpha(\mathbf{R}_i) u_\beta(\mathbf{R}_i)$, as in Ref. [10].

All the parameters of the effective Hamiltonian are computed by performing DFT calculations. To be more specific, the parameters of the second-order terms can be obtained from force constants, while the fourth-order parameters in the self-energy part are estimated by fitting the DFT total energies by the self-energy form. Moreover, the elastic-local mode interacting parameters $B_{l\alpha\beta}$ are obtained by considering the dependence of force constants on strain.

## 2. The spontaneous polarizations and lattice distortions of SnTe thin films

The spontaneous polarizations of SnTe thin films calculated with the Born effective charges (see Fig. S1) as a function film thickness are shown in Fig. S2. For comparison, we also show the average displacements of $Sn^{2+}$ ions. The lattice distortion as a function of film thickness is shown in Fig. S3. We can see that the spontaneous polarizations, the average displacements of $Sn^{2+}$ ions, and the lattice distortions have similar dependence on the film thickness: They first increase, then decreases along with the increase of the film thickness.

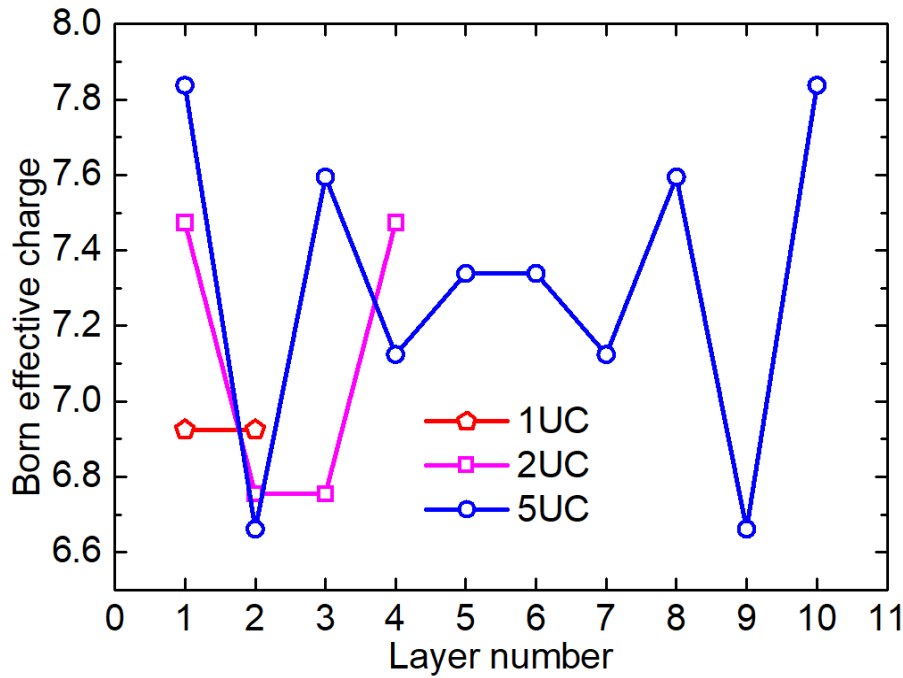

Figure S1. Born effective charges of $Sn^{2+}$ ions in paraelectric SnTe thin films as a function of the layer number. The in-plane component of the Born effective charges is plotted here since it is relevant to the in-plane electric polarization.

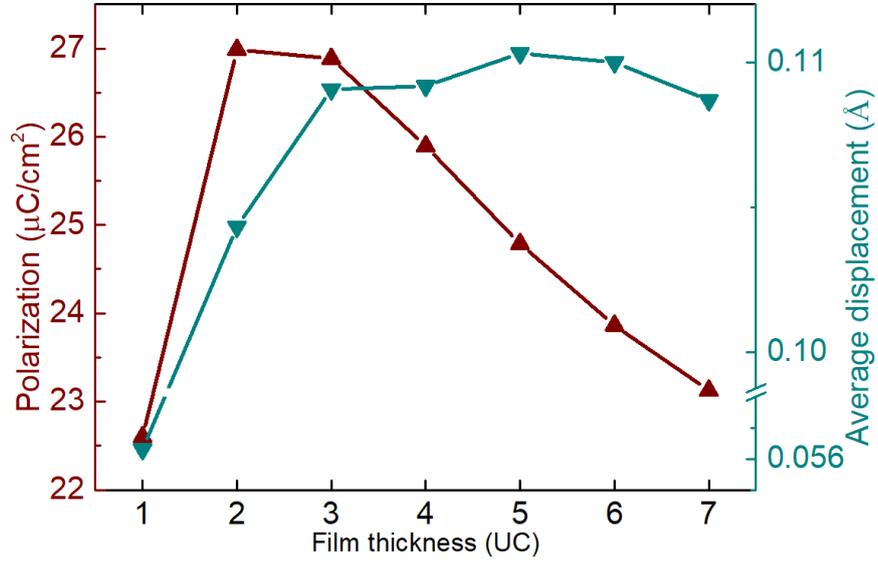

Figure S2. The spontaneous polarization and the average displacements of $Sn^{2+}$ ions of FE SnTe thin films as a function of film thickness. Note that we set the thickness of each SnTe layer to be the average vertical distance between two neighboring SnTe layers when we compute the volume of the SnTe thin films.

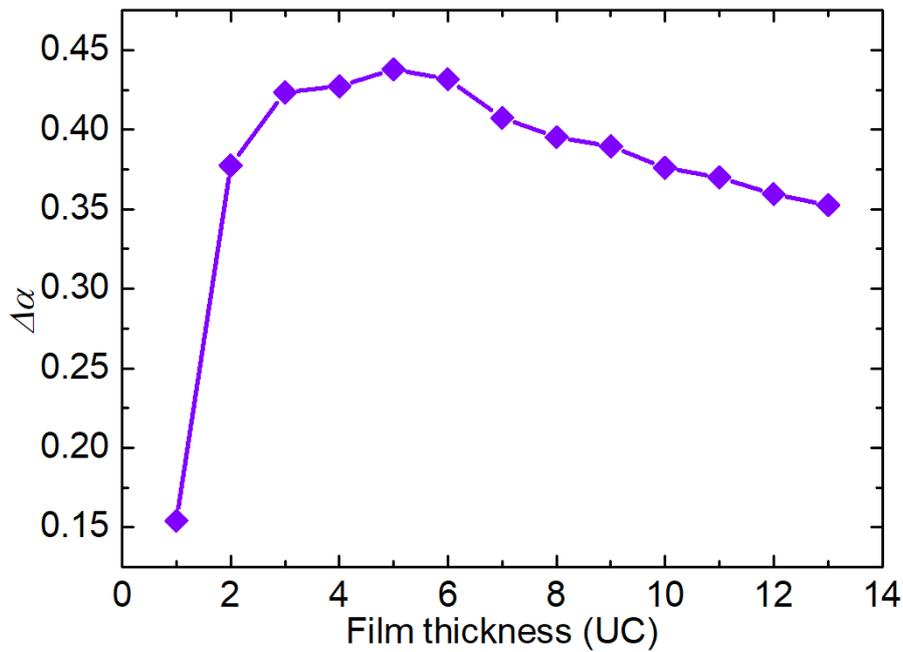

Figure S3. The lattice distortion $\Delta\alpha$ (the distortion angle of the rock-salt unit cell) of FE SnTe thin films as a function of film thickness.

## 3. Layer force constants of $Sn^{2+}$ ions in SnTe thin films as a function of the layer number

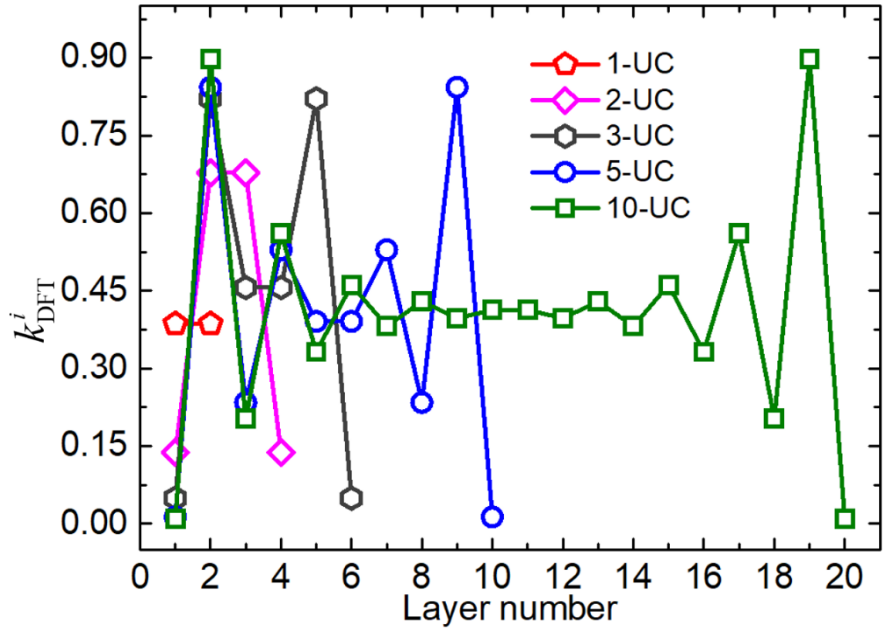

Figure S4. Layer force constants of $Sn^{2+}$ ions as a function of the layer number, as fitted with the DFT total energies.

## 4. Layer force constants of $Mg^{2+}$ ions in MgO thin films

The layer force constants $k_{DFT}^i$ of $Mg^{2+}$ ions in the 5-UC MgO thin film are computed by fitting the total energy $E$ as a function of the in-plane displacement $x$ of the Mg ion with the function $E = \frac{1}{2}k_{DFT}^i x^2 + E_0$ ($E_0$ is the energy of the ground state). Fig. S5 shows the computed layer force constants for the 5-UC MgO thin film. Because MgO is an ionic crystal [12], the layer force constant $k_{DFT}^i$ almost does not contain contribution from the hybridization interaction (HI). Smaller layer force constant of surface $Mg^{2+}$ ion indicates that there is less Pauli repulsion (PR) in a surface layer.

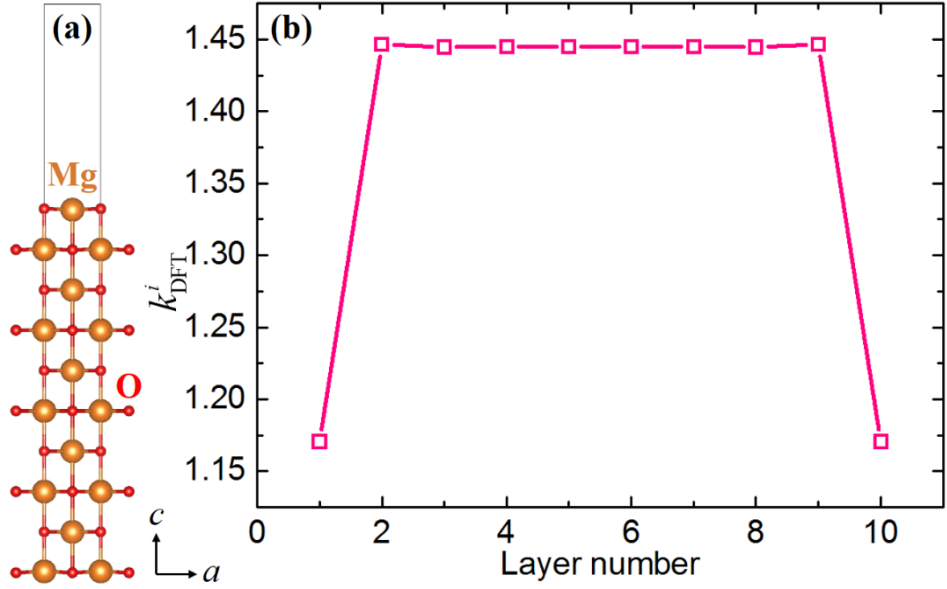

Figure S5. (a) The structure of 5-UC MgO thin film. (b) Layer force constants of $Mg^{2+}$ ions as a function of layer number computed from DFT calculations.

## 5. Energy barriers of GeTe thin films

The FE $T_c$ for bulk GeTe is about 670 K, which is much higher than that for bulk SnTe [13]. Note that the ionicity of GeTe is weaker than that of SnTe, and the covalency of GeTe is stronger than that of SnTe [14], suggesting that the HI in GeTe is stronger than that in SnTe.

For the GeTe thin films, the energy barrier increases with the film thickness, i.e., there is no maximum in the energy barrier curve, as shown in Fig. S7. This can be understood because the HI in GeTe is so strong that the PR contribution to the layer force constant is less important.

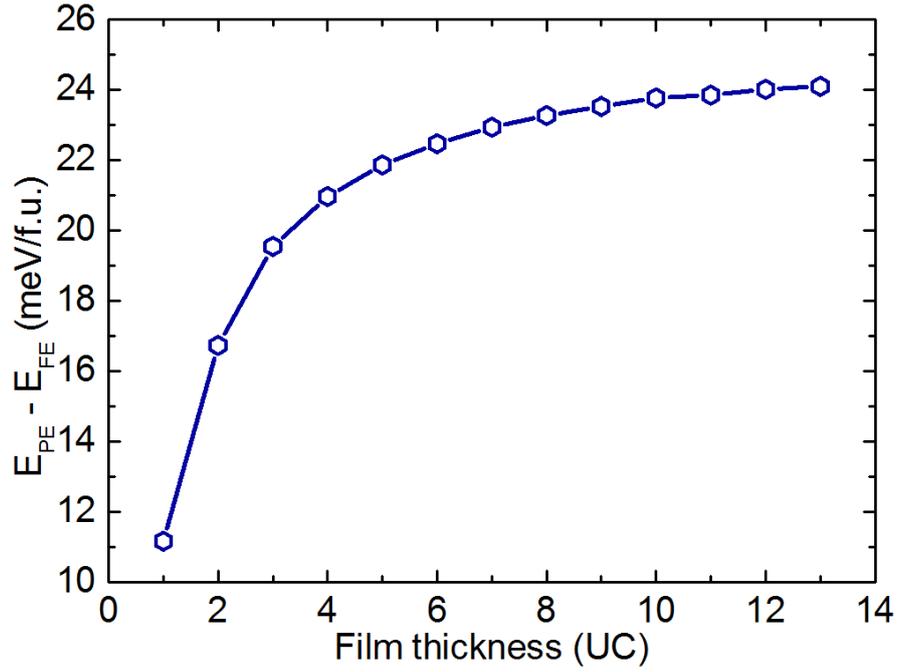

Figure S6. Energy barriers of GeTe thin films as a function film thickness.

**6. Energy barriers of PbTe thin films with tensile strain**

Bulk PbTe is paraelectric [14]. Note that the ionicity of PbTe is stronger than that of SnTe, and the covalency of PbTe is weaker than that of SnTe [15], suggesting that the HI in PbTe is weaker than that in SnTe. We find that the strain-free PbTe thin films remain PE, as expected.

For the PbTe thin films with 2% tensile strain, the energy barrier has the same trend as the SnTe thin films, i.e., it has a maximum for the 5-UC thin film (see Fig. S6). This occurs because the PR and HI in the PbTe thin films with a tensile strain have comparable magnitude.

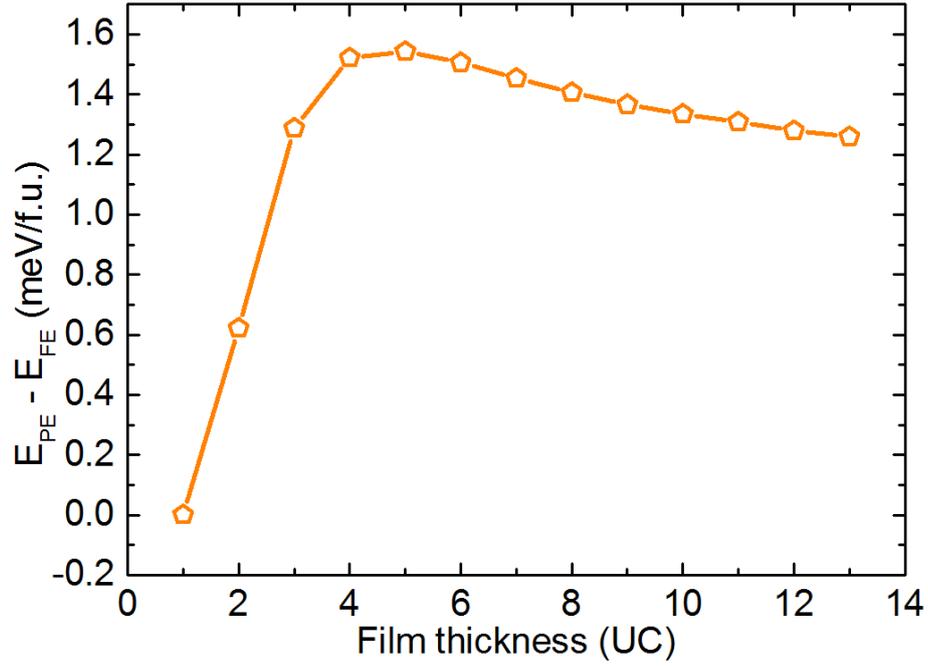

Figure S7. Energy barriers of PbTe thin films with a 2% tensile strain as a function of film thickness.

**7. Energy barriers and polarizations of TiO$_2$-terminated SrTiO$_3$ [001] thin films under a tensile strain (lattice constants $a = b = 4.0955$ Å)**

Interestingly, the enhancement of ferroelectricity in ultrathin films can also take place in other systems. For instance, we find that the energy barriers and polarizations of strained TiO$_2$-terminated SrTiO$_3$ [001] thin films ($a = b = 4.0955$ Å, note that bulk SrTiO$_3$ has an equilibrium lattice constant of 3.905 Å) both increase along with the decrease of the film thickness (see Fig. S8). This is because of the Pauli repulsions in surface layers which is smaller than those in inner layers.

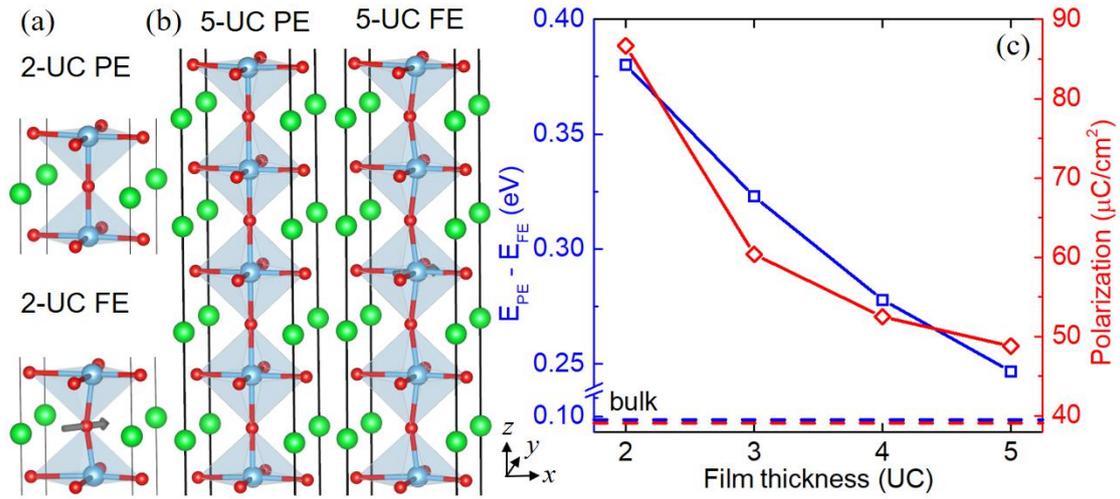

Figure S8. Crystal structures of PE and FE phases of (a) 2-UC and (b) 5-UC strained $TiO_2$-termination $SrTiO_3$ [001] thin films ($a = b = 4.0955$ Å). The gray arrows represent the polarizations along [110] direction. (c) Energy barriers and polarizations as a function of film thickness. Blue and red dash line represent energy barrier and polarization of strained bulk $SrTiO_3$ ($a = b = 4.0955$ Å), respectively.

## 8. Dependence of $T_c$ on film thickness predicted from PTMC simulations using an effective Hamiltonian

We first construct an effective Hamiltonian by performing DFT calculations with the PBE functional and including the 4d electrons of the Sn ions as valence electrons (i.e., the "Sn_d" procedure). Then we perform PTMC simulations to find the ferroelectric transition temperature $T_c$ of SnTe systems. Our results show that $T_c$ increases from 1-UC to 5-UC and then decreases when further increasing the film thickness, as shown in Fig. S9. SnTe thin-films have higher $T_c$ than bulk SnTe except for the 1-UC SnTe thin-film case. The tendency of $T_c$ with thickness is in line with that of the energy barrier.

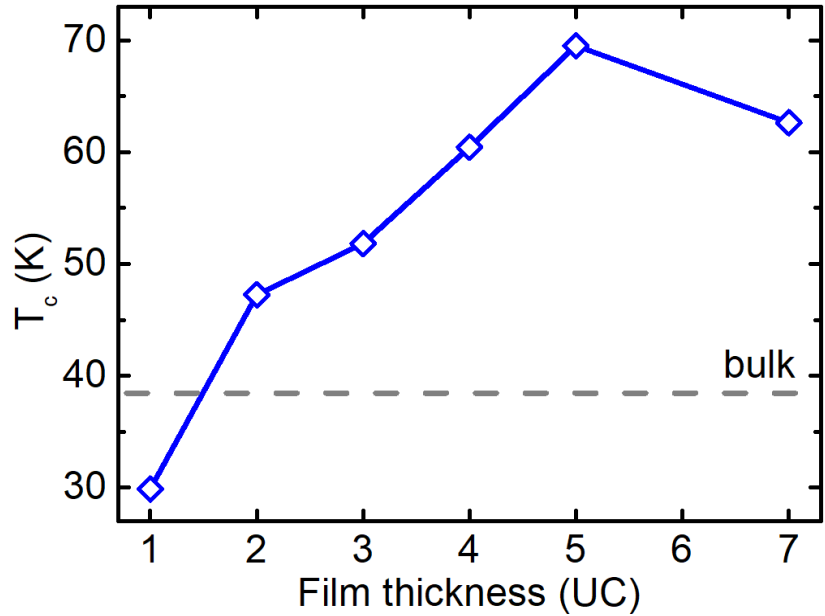

Figure S9. Ferroelectric transition temperature $T_c$ of SnTe thin films as a function of thickness, as predicted from PTMC simulations using an effective Hamiltonian. The dashed line corresponds to the theoretical $T_c$ of bulk SnTe.

## 9. FE switching energy barriers in SnTe systems calculated by different methods

Fig. S9 shows that the FE Curie temperature of bulk SnTe and SnTe thin films estimated by using a first-principles-based effective Hamiltonian are underestimated. This is mainly due to the fact that our DFT calculations underestimate the FE switching energy barriers in SnTe systems. In our DFT calculations for extracting the parameters of the effective Hamiltonian, we adopt the PBE functional and include the 4d electrons of the Sn ions as valence electrons (this computing procedure is referred as "Sn_d"). However, we find that a more accurate hybrid functional (i.e., HSE) and/or the spin-orbit coupling effect will enhance the FE switching energy barriers in SnTe systems. In fact, these more accurate methods predict higher energy barriers for bulk SnTe in most cases (see Fig. S10). In particular, the energy barrier in bulk SnTe calculated by the accurate "HSE+SOC+Sn_d" procedure is four times as large as that by the "Sn_d" procedure. Since a larger energy barrier usually suggests a higher FE $T_c$, we expect that the real $T_c$ will be enhanced by four times if we extract the parameters of the effective Hamiltonian with the "HSE+SOC+Sn_d" procedure.

In the present work, we extracted the parameters of the effective Hamiltonian with the "Sn_d" procedure since the more accurate "HSE+SOC+Sn_d" procedure is much more demanding. However, our test calculations show that the more accurate procedures give similar physical trend. For example, the energy barriers of SnTe thin films computed with the "Sn_d+SOC" procedure first increases with thickness, but then decreases for thicker films (shown in Fig. S11). This result is in qualitative agreement with the result from the "Sn_d" procedure (quantitatively, the "Sn_d+SOC" procedure predicts a larger energy barrier).

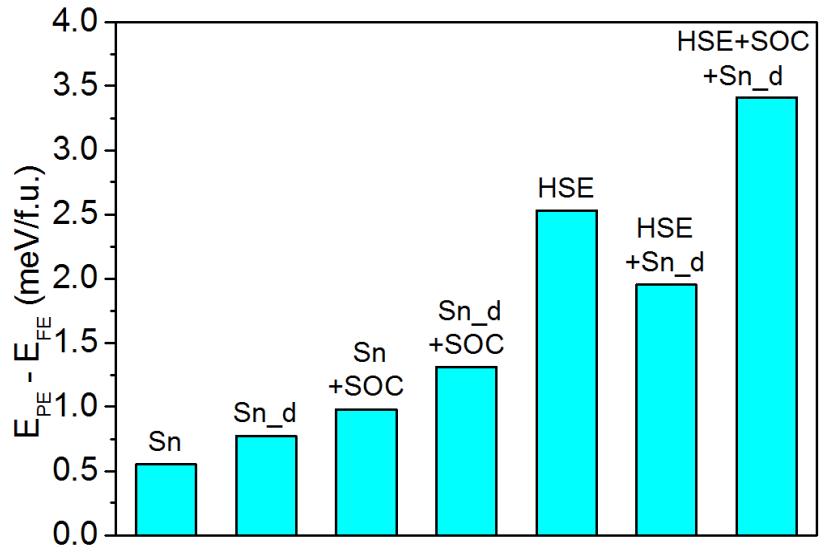

Figure S10. Energy barriers of bulk SnTe, as calculated by using different methods. In the "Sn" procedure, the 4d electrons of the Sn atom are treated as core electrons. In contrast, "Sn_d" means that the ten 4d electrons of the Sn atom are taken as valence electrons. "SOC" indicates that spin-orbit coupling is included in the calculation. "HSE" means that the hybrid HSE06 exchange-correlation functional is adopted.

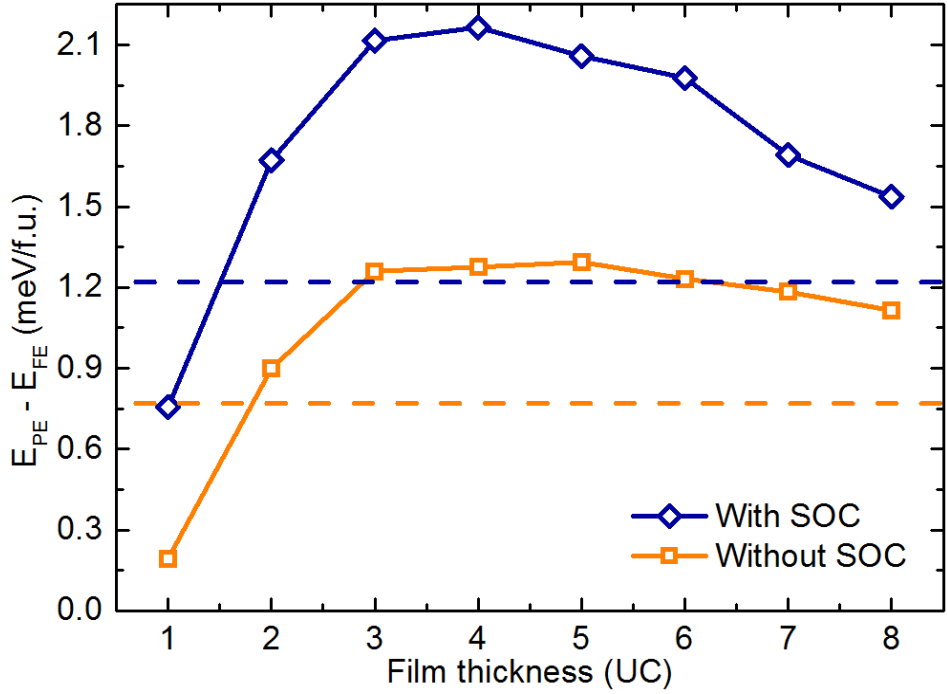

Figure S11. Energy barriers of SnTe thin films as a function of thickness for the case of "Sn_d" and "Sn_d+SOC", respectively. The orange and blue horizontal dashed lines represent energy barriers of bulk SnTe for the case of "Sn_d" and "Sn_d+SOC", respectively.

**10. Energy barriers of SnTe systems with Sn vacancy**

Energy barriers of SnTe systems with 6.25% Sn vacancy are listed in Table I. For comparison, we also report the energy barriers of SnTe systems without vacancy. Our results show that Sn vacancies suppress ferroelectric stability in SnTe systems. Note also that the energy barrier of 1-UC SnTe thin film without vacancies is higher than that of bulk SnTe with vacancies, which is in agreement with an extrinsic explanation proposed in Ref. [16] about the enhancement of the Curie temperature when going from bulk to 1-UC SnTe film.

Table I. Energy barriers of SnTe systems with/without Sn vacancy (in meV).

|  | 1-UC | 2-UC | bulk |
| --- | --- | --- | --- |
| With vacancies | 0.034 | 0.142 | 0.027 |
| Without vacancies | 0.195 | 0.900 | 0.340 |

## 11. Dependence of $T_c$ on film thickness from the simulations with a simple model

Above we used a realistic effective Hamiltonian to estimate the $T_c$ Curie temperature in SnTe thin films. In order to demonstrate more clearly the important role of the force constant of the surface ions on the dependence of $T_c$ on film thickness, we adopt here a simpler model. More precisely, we consider a simple bulk structure where the dipoles form a tetragonal lattice [shown in Fig. S12(a)], and the FE polarization is along the in-plane [110] direction. For the bulk structure, this simple Hamiltonian can be written as:

$$E^{tot} = E^{self}(\{\mathbf{u}\}) + E^{short}(\{\mathbf{u}\}),$$

where $E^{self}(\{\mathbf{u}\}) = \sum_i [k_2 u^2 + k_4(u_{ix}^4 + u_{iy}^4) + k_4' u_{ix}^2 u_{iy}^2]$ (where, for simplicity, we set $k_4 = k_4' > 0$). The short-range interaction $E^{short}(\{\mathbf{u}\}) = \sum_{<i,j>} J(u_{ix} u_{jx} + u_{iy} u_{jy})$, where we used the same interaction parameter for all nearest neighboring interactions (both in-plane and out-of-plane). For the [001] thin films [The atomic model of the 3-layer thin film is shown in Fig. S12(b)], the second-order force constant of the surface dipoles can be different from that of the inner dipoles. To be more specific, the self-energy for the thin film can be written as: $E^{self}(\{\mathbf{u}\}) = \sum_{i \in surface} k_2^{surface} u^2 + \sum_{i \in inner} k_2^{inner} u^2 + \sum_i [k_4(u_{ix}^4 + u_{iy}^4) + k_4' u_{ix}^2 u_{iy}^2]$, where $k_2^{inner}$ is the same as that in the bulk structure (i.e., $k_2^{inner} = k_2$).

We used this simple Hamiltonian to estimate $T_c$ of thin films as a function of film thickness, as shown in Figs. S12(c) and (d). When $k_2^{surface} = k_2$, $T_c$ increases with the film thickness, in agreement with previous beliefs [17-20]. When $k_2^{surface} < k_2$, $T_c$ displays the same trend as in Fig. S9. This proves that the unusual phenomenon that $T_c$ has a maximum at a certain thickness is due to the smaller force constants of surface dipoles.

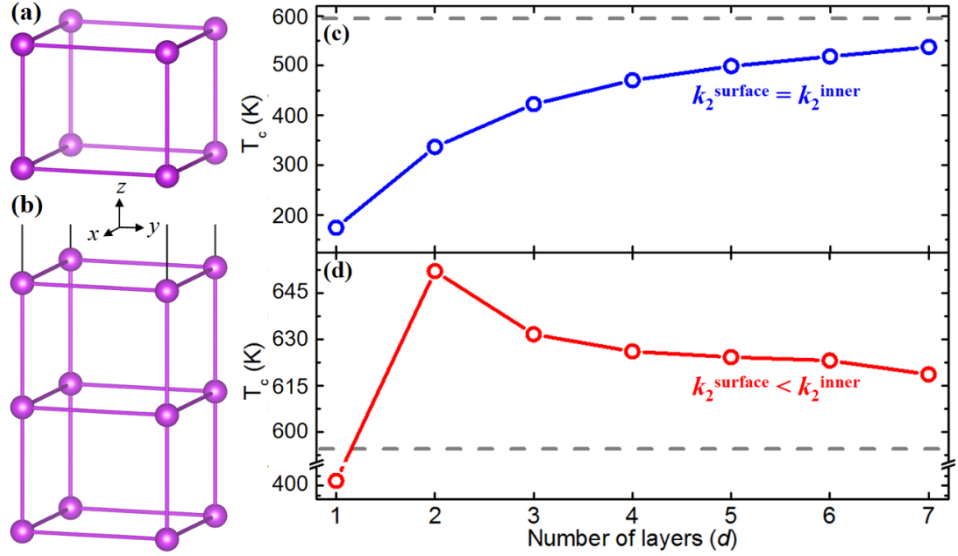

Figure S12. $T_c$ as a function of film thickness estimated with the simple Hamiltonian discussed in the Supplemental Material. The bulk structure and 3-layer thin film structure are shown in (a) and (b), respectively. The results for $k_2^{surface} = k_2 = 0.05$ and $k_2^{surface} = -0.1 < k_2 = 0.05$ are shown in Panels (c) and (d), respectively. The other parameters are $k_4 = 0.4$ and $J = -0.1$. The horizontal dashed line represents the $T_c$ for the bulk structure.